\input{aipcheck}

\documentclass[
    ,final            
  ]
  {aipproc}

\layoutstyle{6x9}

\begin{document}

\title{Probing the Low Mass X-ray Binaries/Globular Cluster connection in NGC1399}


\author{M. Paolillo}{address={Department of Physical Sciences, University Federico II, Naples, Italy}}

\author{T. Puzia}{address={Herzberg Institute of Astrophysics, Victoria, BC}}

\author{S. E. Zepf }{address={Department of Physics and Astronomy, Michigan State University, USA}}
\author{T.J. Maccarone}{address={School of Physics and Astronomy, University of Southampton, UK}}
\author{P. Goudfrooij}{address={Space Telescope Science Institute, Baltimore, MD 21218, USA}}
\author{A. Kundu}{address={Eureka Scientific, Oakland, CA 94602, USA}}
\author{G. Fabbiano}{address={Harvard-Smithsonian Center for Astrophysics, Cambridge, USA}}
\author{L. Angelini}{address={Laboratory for High Energy Astrophysics, NASA Goddard Space Flight Center, USA}}

\begin{abstract}
We present a wide field study of the Globular Clusters/Low Mass X-ray Binaries connection in the cD elliptical NGC1399, combining HST/ACS and {\it Chandra} high resolution data. We find evidence that LMXB formation likelihood is influenced by GCs structural parameters, in addition to the well known effects of mass and metallicity, independently from galactocentric distance.
\end{abstract}
\maketitle


\vspace{-0.8cm}\section{Optical  and X-ray Data}\vspace{-0.2cm}
The optical data were taken with the HST/ACS camera (F606W filter), arranged in a 3x3 mosaic covering ~100 sq.arcmin and extending out to ~55 kpc, i.e. 1.6 $r_e^{GC}$ (~5.7 $r_e^{gal}$). Source catalogs were generated with SExtractor, and registered to the USNO-B1 reference frame (final accuracy of 0.2"). 
Structural parameters were measured fitting King models to the data with the Galfit code \cite{Peng02}, deriving tidal, core, effective radii and central surface brightness; our accuracy was estimated simulating artificial GCs with the Multiking code (\url{http://people.na.infn.it/paolillo/Software.html}) which accounts for field distortion, PSF variation, dithering pattern.
The X-ray data have been retrieved from the Chandra archive (obsid 319 and 1472) for a total exposure time of ~100 ks. 230 X-ray sources were detected in the 0.3-1, 1-2, 2-8 keV bands; astrometric and photometric measurements were derived with the ACIS Extract software \cite{Broos02} obtaining a residual positional uncertainty of 0.3" r.m.s.  Adopting a 1" matching radius results in 138 GC-LMXB candidates.
\vspace{-0.4cm}

\section{LMXB Properties and Formation Scenarios}\vspace{-0.2cm}
We confirm that NGC 1399 has an unusually large fraction (65\%) of LMXB residing in GCs even after taking into account its high GC specific frequency \cite[$S_n$;][]{Angelini01, Kim04, Kundu07}. Our data indicate that this galaxy is inconsistent at $3\sigma$ level with the 50\% value found for galaxies with similar $S_n$ (e.g. NGC 4649 or NGC 4472). This value however depends on galactocentric distance ranging from 50\% for r<50" to >75\% at r>200".
\begin{figure}
  \includegraphics[height=.35\textheight]{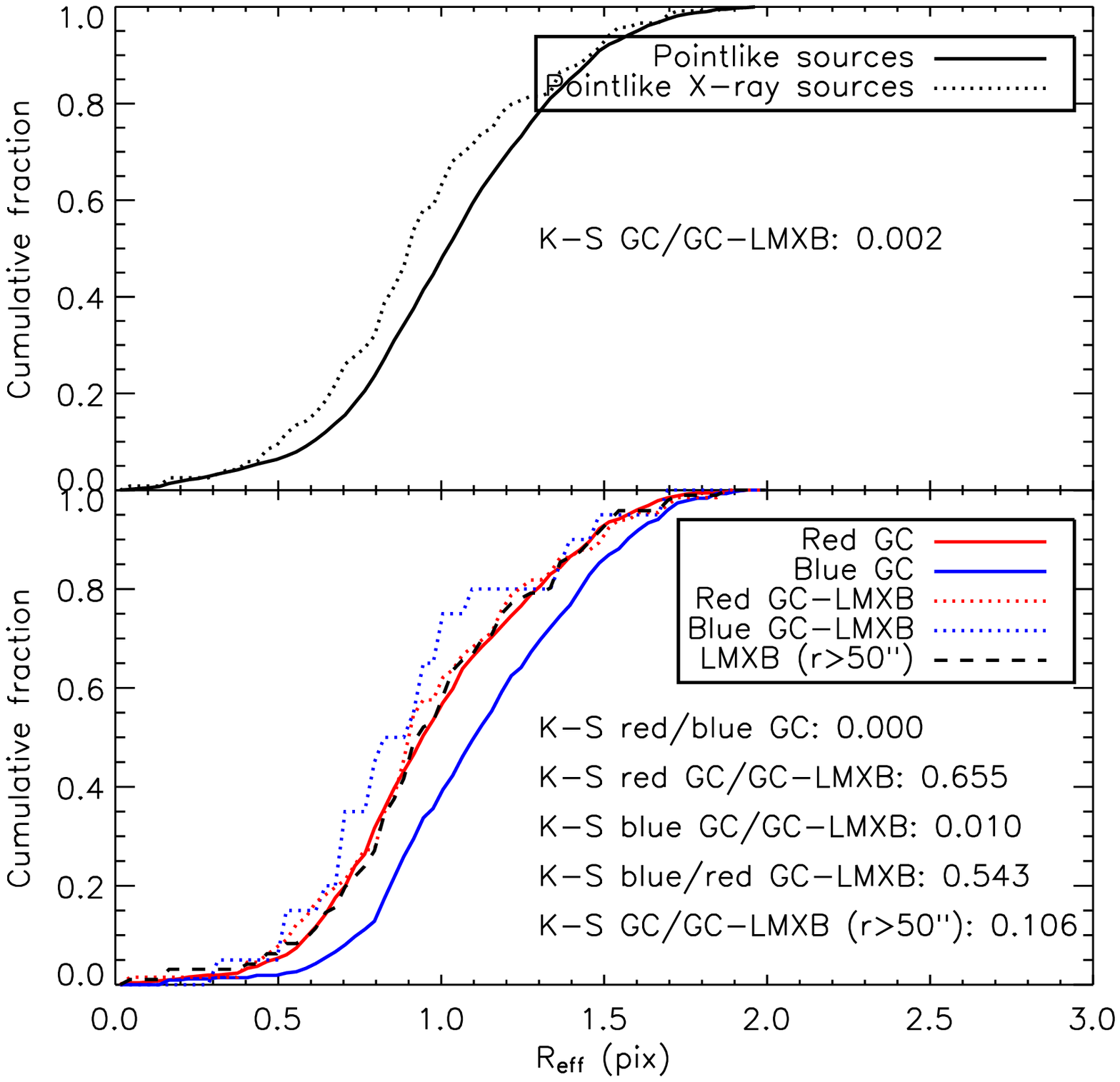}
  \includegraphics[height=.355\textheight]{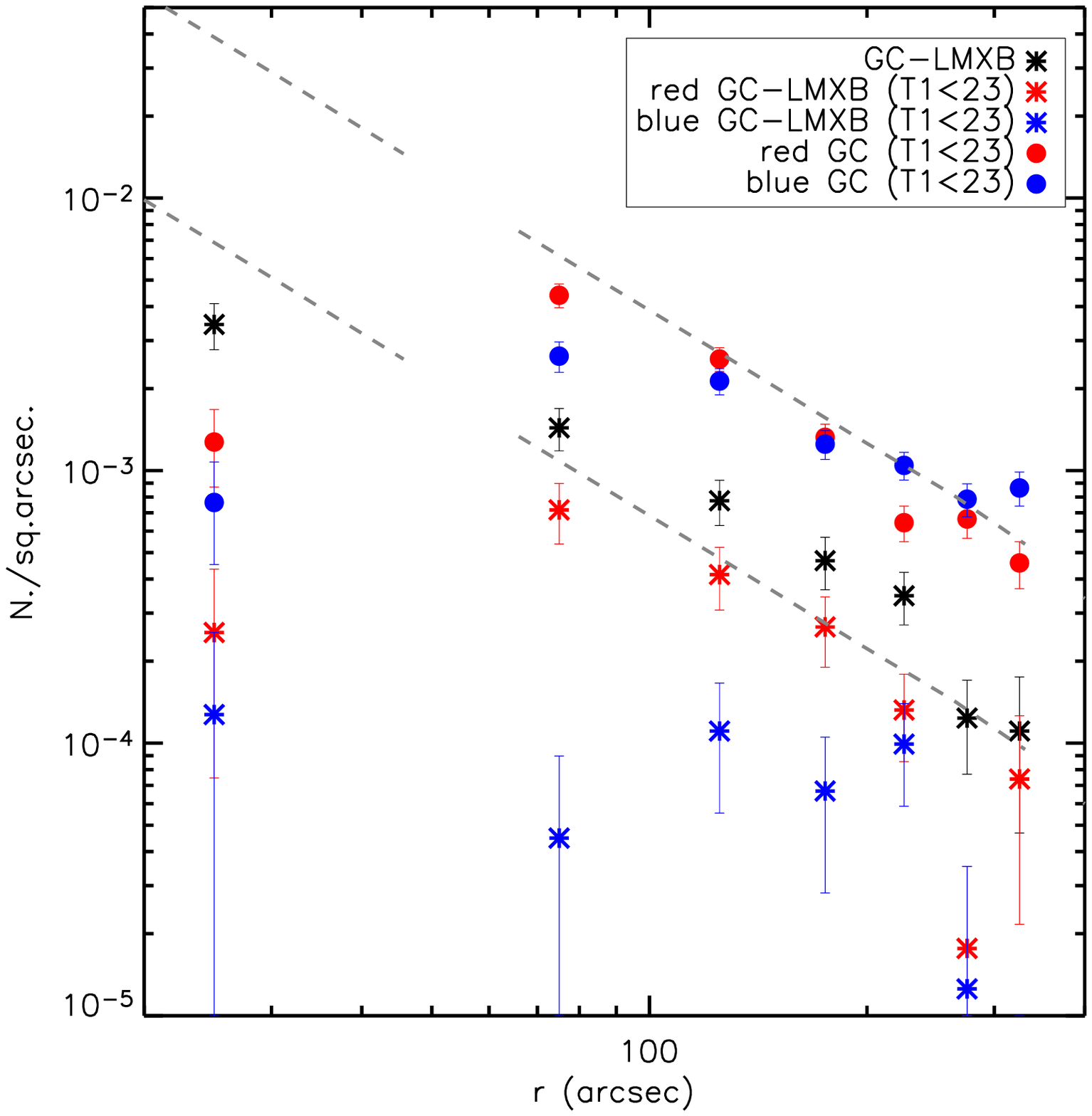}
  \caption{{\it Left:} Cumulative distributions and K-S tests comparing effective radii of different GC populations. {\it Right:} Radial surface density distributions of GCs and LMXBs.\vspace{-0.7cm}}
\end{figure}

X-ray sources tend to reside in bright GCs, in agreement with the expectation that LMXB formation likelihood is related to GC mass. On average, ~6\% of GCs with $m_V<26$ host LMXBs, but the actual fraction depends on the magnitude and color of the host GC: LMXB are preferentially hosted by bright ($m_V<24$) and red ($C-R>1.65$) GC with fractions that can be >20\%. Red GCs are $>3$ times more likely to host a LMXB than blue GCs, but this ratio also depends on magnitude and, to a lesser extent, on galactocentric distance (Fig.1).
NGC1399 also has a very red GC sub-population hosting the majority of LMXB in the galaxy center \cite{Kundu07}, which disappears for $r>100"$.

As already found by other authors there is no obvious difference in the properties (luminosity function, X-ray color) of field, red and blue LMXBs, with the only exception that the brightest LMXBs seem to reside in GCs. The detection of X-ray variability in 1/3 of the bright LMXBs ($L_X>3\times 10^{38}$ erg/s) argues against the superposition of several X-ray sources within the same host GC.

We do not detect any significant difference between the GC-LMXB radial distribution and the overall GC distribution (Fig.1). Our data show that LMXBs hosted by red GC simply follow their parent distribution (the statistics is too low for blue LMXBs to draw significant conclusions). 
The radial distribution of field LMXB is significantly steeper than GC-LMXBs, at the >99.9\% level. Field LMXBs are thus unlikely to be originally formed in GCs and later ejected from the host GC. Field LMXBs in fact follow closely the surface brightness of the host galaxy, suggesting that they are formed within the main stellar body.
LMXB GCs are on average more compact than the bulk of the population, but do not differ significantly from the red GC population which hosts the majority of them (Fig.1). While red and blue GC have different sizes at >99\% level, we cannot detect any significant difference between red LMXB GCs and the overall red GC population.  On the other hand LMXB residing in blue GC prefer the most compact systems.

\end{document}